\documentclass[twocolumn,aps,showpacs,prb]{revtex4-1}
\usepackage{graphicx}
\usepackage{subfigure}
\usepackage{epstopdf}
\usepackage{amsmath}
\usepackage{multirow}
\usepackage{tabularx}
\usepackage{color,ulem}
\begin{document}
\title{Coexistence of topological Weyl and nodal-ring states in ferromagnetic and ferrimagnetic double perovskites}
\author{Xinlei Zhao$^{1}$}
\author{Peng-jie Guo$^{2}$}
\author{Fengjie Ma$^{1}$}\email{fengjie.ma@bnu.edu.cn}
\author{Zhong-Yi Lu$^{3}$}

\date{\today}

\affiliation{$^{1}$The Center for Advanced Quantum Studies and Department of Physics, Beijing Normal University, 100875 Beijing, China}
\affiliation{$^{2}$Songshan Lake Materials Laboratory, Dongguan, Guangdong 523808, China}
\affiliation{$^{3}$Department of Physics, Renmin University of China, Beijing 100872, China}

\begin{abstract}
Magnetic topological quantum materials have attracted great attention due to their exotic topological quantum physics induced by the interplay among crystalology, magnetism, and topology, which is of profound importance to fundamental research and technology applications. However, limited materials are experimentally available, most of whom are realized by magnetic impurity doping or heterostructural constructions. In this work, based on the first-principles calculations, we predict that double perovskite Ba$_2$CdReO$_6$ is an intrinsic ferromagnetic topological semi-half-metal, while the ferrimagnetic double perovskite with space group symmetry $Fm$$\bar{3}$$m$, such as Ba$_2$FeMoO$_6$, belongs to a topological half-metal. One pair of Weyl points and fully spin-polarized nodal-ring states are found in the vicinity of the Fermi level in Ba$_2$CdReO$_6$. Its two-dimensional nearly flat drumhead surface states are fully spin-polarized. In Ba$_2$FeMoO$_6$, however, there exist four pairs of Weyl points and two fully spin-polarized nodal-rings near the Fermi level. These topological properties are stable in the presence of spin-orbit coupling. This makes these materials be an appropriate platform for studying the emerging intriguing properties, especially for the applications in spintronics, information technology, and topological superconductivity.
\end{abstract}

\maketitle

\section{Introduction}

Topological materials have great potential in many frontier fields due to the existence of nontrivial topological quantum states and have become a hot topic in materials science and condensed matter physics \cite{RevModPhys.90.015001, Gao10.1146, vergniory2019complete, tang2019comprehensive, zhang2019catalogue, xu2015observation, liang2015ultrahigh, feng2015large, huang2015observation, zhang2016signatures, weng2015weyl, PhysRevB.97.241102, zhang2017dirac, PhysRevB.96.045121, Huang1180}. Recently, the discovery of time-reversal breaking magnetic topological materials that have exotic properties beyond non-magnetic topological systems, opens a new burgeoning field and attracts more and more attention \cite{zou2019study, nie2017topological, chang2016room, zhang2020nodal, PhysRevLett.124.076403, wang2016time, yang2017topological, tang2016dirac, li2018topological, AFMTI,zhang2019topological, xu2011chern, liu2018giant, Sanchez2020, PhysRevB.99.165117, Li2020, PhysRevX.8.041045, PhysRevB.85.012405, K_bler_2016, Belopolski1278, wang2018large, gong2019experimental, kim2018large, kuroda2017evidence, xue2013qah, wan2011topological, fang2016topological, weng2015topological, you2019two, PhysRevLett.122.057205, XuYCoC2APL, PhysRevB.99.035125, PhysRevB.98.121103, wu2019weyl, PhysRevMaterials.3.021201, he2019topological}. Magnetic topological materials have many advantages due to the interplay between magnetism and topology that offers an imaginative degree for exploring the emerging topological quantum physics, which is of profound importance to fundamental research and technology applications. For example, topological axion states were predicted to exist at the surface of a three-dimensional antiferromagnetic (AFM) topological insulators \cite{AFMTI,zhang2019topological}; quantized anomalous Hall effect can be achieved in intrinsic ferromagnetic (FM) Weyl semimetals without an external magnetic field \cite{xu2011chern,wang2018large, liu2018giant,Sanchez2020,PhysRevB.99.165117,Li2020,PhysRevX.8.041045,PhysRevB.85.012405,K_bler_2016,Belopolski1278}; and the emergence of Majorana modes with non-Abelian statistics in the edge/corner states of magnetic topological materials is closely associated with topological superconductivity and relevant for topological quantum computations \cite{li2018topological,RevModPhys.83.1057}. Great effort has been devoted to the search of new magnetic topological quantum materials. However, only very few limited materials are experimentally available, most of whom are realized by magnetic impurity doping or heterostructural constructions that are quite challenging for realistic fabrication \cite{gong2019experimental, kim2018large, wang2018large, kuroda2017evidence, xue2013qah}.

Magnetic topological semimetals are of particular importance in magnetic topological materials, since quantum transport, spin dynamics, magnetism, as well as topology are all correlated in these systems. The spin-polarized valence and conduction bands of magnetic topological semimetals are linearly intersected due to nontrivial topology, forming zero-dimensional discrete points or one-dimensional continuous curves with nontrivial topological numbers, especially those around the Fermi level \cite{wan2011topological, fang2016topological}. In addition to the advantages of regular topological materials, the high mobility carriers with readily manipulative spins through external fields and disentangled distinctive topological states near the Fermi level may give rise to exotic transport and spectroscopic behaviors. For example, large anomalous Hall conductivity and giant anomalous Hall angle were reported in ferromagnetic kagome-lattice Weyl semimetal Co$_3$Sn$_2$S$_2$, non-centrosymmetric PrAlGe, and several Heusler alloys, which are more than a half order of magnitude larger than those in typical magnetic systems \cite{wang2018large, liu2018giant, Sanchez2020,PhysRevB.99.165117,Li2020,PhysRevX.8.041045,PhysRevB.85.012405,K_bler_2016,Belopolski1278}.

Recently, topological nodal-line spin-gapless semi-metals (TNLSGSMs) were proposed as a new type of topological quantum materials with coexistence of spin fully polarized and linearly dispersive nodal-line states in the vicinity of the Fermi level \cite{zhang2020nodal}. Two-dimensional drumhead surface states are considered to be the fingerprint of topological nodal-line semi-metals. They usually lie in the mirror plane of the Brillouin zone (BZ) protected by the mirror symmetry and produce a large surface density of states \cite{fang2016topological, weng2015topological}. The two-dimensional drumhead surface states in TNLSGSMs are fully spin polarized, which would be greatly beneficial to spintronics and equal spin-pairing topological superconductivity \cite{zhang2020nodal}. However, TNLSGSMs are rarely realized experimentally. One main reason is that FM semi-metals are very rare in nature, regardless of its topological trivial/nontrivial character. In addition, the nodal line in semi-metals is normally deformed into discrete points or fully gapped in the presence of spin-orbit coupling (SOC). There are very few limited materials predicted to be TNLSGSMs that are stable against the SOC \cite{xu2011chern,zhang2020nodal}.

Double perovskites (DPs), which have a general chemical formula of A$_2$BB$'$O$_6$, represent a rich playground in the field of materials research. The A-site cation normally is alkaline earth or rare earth element, where B/B$'$ adopts a cation of transition metal, alkali metal, alkaline earth, or main group metal element. Due to the flexible combination of magnetic or non-magnetic B and B$'$ ions, the DP A$_2$BB$'$O$_6$ family exhibits various crystal structures, rich electronic properties, and abundant magnetic structures. To be specific, half-metallic DPs such as Sr$_2$FeMoO$_6$ exhibit intrinsic tunneling-type magnetoresistance even at room temperature, whose spin fully-polarized properties around the Fermi level are also very useful for spintronic applications \cite{Kobayashi1998, saloaro2016toward}. Superconductivity has been found in a number of Ru-based DPs by partially substituting the Ru ion with the Cu ion \cite{wu1996superconductivity,DPSC}. There are many other fascinating properties of DPs that enthrall researchers, such as photovoltaic \cite{roknuzzaman2019electronic}, multiferroic \cite{ma4010153}, thermoelectric \cite{sugahara2008thermoelectric}, and electronic transport \cite{cook2014double}. The extreme flexibility of DPs in terms of symmetry and degrees involved enriches the exploration of other prospective properties.

In this paper, we reveal the rich and interesting topological properties in two types of DP compounds with space group symmetry $Fm$$\bar{3}$$m$. By first-principles calculations, we predict that Ba$_2$CdReO$_6$ is a new type of intrinsic FM TNLSGSM. The barium based compound Ba$_2$CdReO$_6$ has been experimentally synthesized \cite{Sleight1962Compounds}. However, there is no experimental study about its magnetism so far. Our calculations presented here predict that Ba$_2$CdReO$_6$ has an FM ground state. Moreover, it is an ideal intrinsic FM TNLSGSM that possesses a pair of Weyl points and one nodal-ring in the first BZ. In addition, we have also studied the topological aspects of ferrimagnetic (FiM) half-metals of DPs, which have both B and B$'$ transition metal ions. Ferrimagnetic half-metals are another very interesting class of DPs that have been widely studied \cite{AFMO_Fim, AFRO_Fim, ACWO_Fim, PFMO_Fim}. They generally show extremely high Curie temperatures above room temperature and great low-field tunneling magnetoresistance, which are highly desirable for promising potential applications in spin electronics \cite{AFMO_Fim, AFRO_Fim, ACWO_Fim, PFMO_Fim}. We find that FiM Ba$_2$FeMoO$_6$ is a topological half-metal, which has four pairs of Weyl points and two fully spin-polarized nodal rings in the first BZ. Other similar structural DP compounds, such as Ba$_2$FeReO$_6$, Sr$_2$CrWO$_6$, and Pb$_2$FeMoO$_6$, share similar topological characters, and all belong to FiM topological half-metals.

\section{Computational details}

In our calculations, the plane-wave basis based method and Quantum-ESPRESSO software package were used \cite{QE2009, QE2017}. We adopted the generalized gradient approximation (GGA) of Perdew-Burke-Ernzerhof formula for the exchange-correlation potentials in the electronic structure simulations \cite{perdew1996generalized}. The ultrasoft pseudopotentials were employed to model the electron-ion interactions \cite{vanderbilt1990soft}. A corrective Hubbard-like $U$ term was introduced to treat the strong on-site Coulomb interaction of the localized electrons of the transition metal ions \cite{PhysRevB.52.R5467, PhysRevB.71.035105}. The effective values of $U$ used in calculations were 3.0, 4.3, and 2.4 eV for Re-$5d$, Fe-$3d$, and Mo-$4d$ electrons, respectively. \cite{PhysRevB.52.R5467, PhysRevB.71.035105}.
Parameter-free \textit{ab initio} calculations with the recently developed strongly constrained and appropriately normed (SCAN) meta-GGA exchange-correlation functional were also performed for crossing checks, which had been demonstrated to be able to give an accurate descriptions of a number of strongly correlated materials without invoking any free parameters such as the Hubbard $U$ \cite{SCANLCO, SCANYBCO}. A mesh of \mbox{16$\times$16$\times$16} \mbox{k-points} grid was used for sampling the BZ, and the marzari-vanderbilt broadening technique was adopted \cite{marzari1999thermal}. After full convergence tests, the kinetic energy cutoff for wavefunctions and charge densities were chosen to be 80 and 640 Ry, respectively. During the simulations, all structural geometries were fully optimized to achieve the minimum energy. The surface states were studied using tight-binding methods by the combination of Wannier90 \cite{mostofi2008wannier90} and WannierTools \cite{WU2017} software packages.

\section{Results and discussion}

\subsection{TNLSGSM of Ba$_2$CdReO$_6$}


\begin{figure}
\centering
\includegraphics[width=8.5cm]{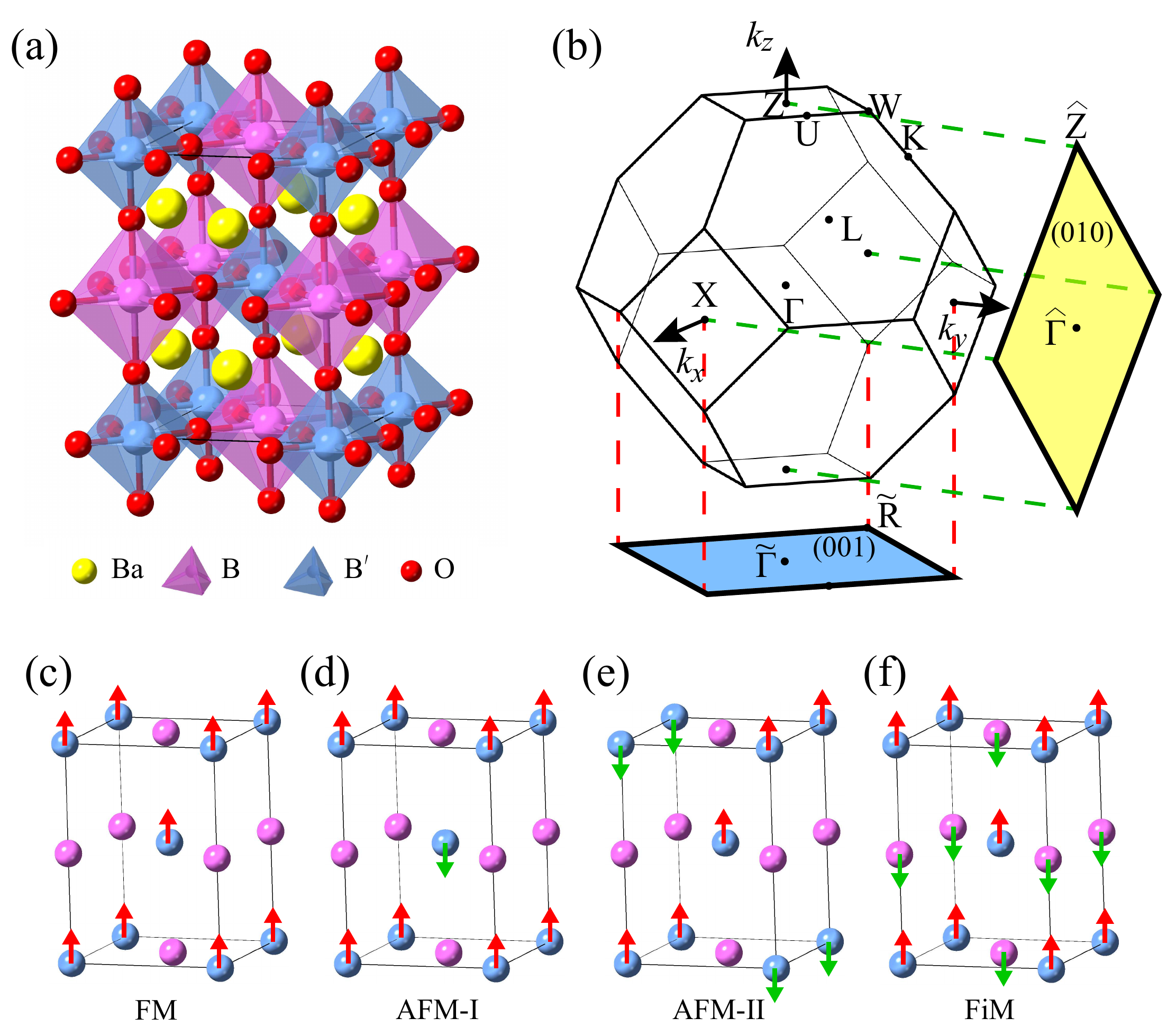}
\caption{\label{crystal} (a) Crystal structure of DP A$_2$BB$'$O$_6$ with $Fm$$\bar{3}$$m$ space group symmetry. (b) The bulk Brillouin zone (BZ) with the high symmetry points labeled. The projected surface BZs of (001) and (010) planes are depicted by blue and yellow semitransparent planes, respectively. (c-f) The different magnetic configurations used for the ground state calculations.}
\end{figure}

The crystal structure of DP Ba$_2$CdReO$_6$ belongs to an ideal cubic rock-salt type with space group symmetry $Fm$$\bar{3}$$m$ (No. 225), as shown in the Fig. \ref{crystal}(a). Its unit cell is doubled in comparison with that of a single perovskite \cite{green2014the}. The corresponding first BZ with high symmetry points is shown in the Fig. \ref{crystal}(b). In our calculations, several possible magnetic orders of ground states were considered, including AFM-I, AFM-II, FM, and non-magnetic (NM) states, as shown in the Figs. 1(c)-1(e) \cite{PhysRevB.99.104411}. The spins on Re atoms alternate by layer along the $c$ axis for the AFM-I state, while [101] direction for the AFM-II state. The calculations reveal that DP Ba$_2$CdReO$_6$ has an FM ground state. If we set the energy of FM Ba$_2$CdReO$_6$ be zero as a reference, the relative energies of AFM-I, AFM-II, and NM states with respect to the FM state are 6.4, 19.0, and 120.3 meV per formula unit cell, respectively. After full relaxation of geometries, the optimized lattice constant is a=8.41$\AA$, in good agreement with the experimental value \cite{Sleight1962Compounds}. The ordering magnetic moment is mainly around Re ion with electronic configuration of Re$^{6+}$ (5d$^{1}$), $\sim$ 1.0 $\mu_{B}$. Similar results can be obtained using more advanced SCAN meta-GGA functional. Therefore, with the selected $U$ value, our GGA+$U$ calculations can describe the electronic structures of this material very well.


\begin{figure}
\centering
\includegraphics[width=8.5cm]{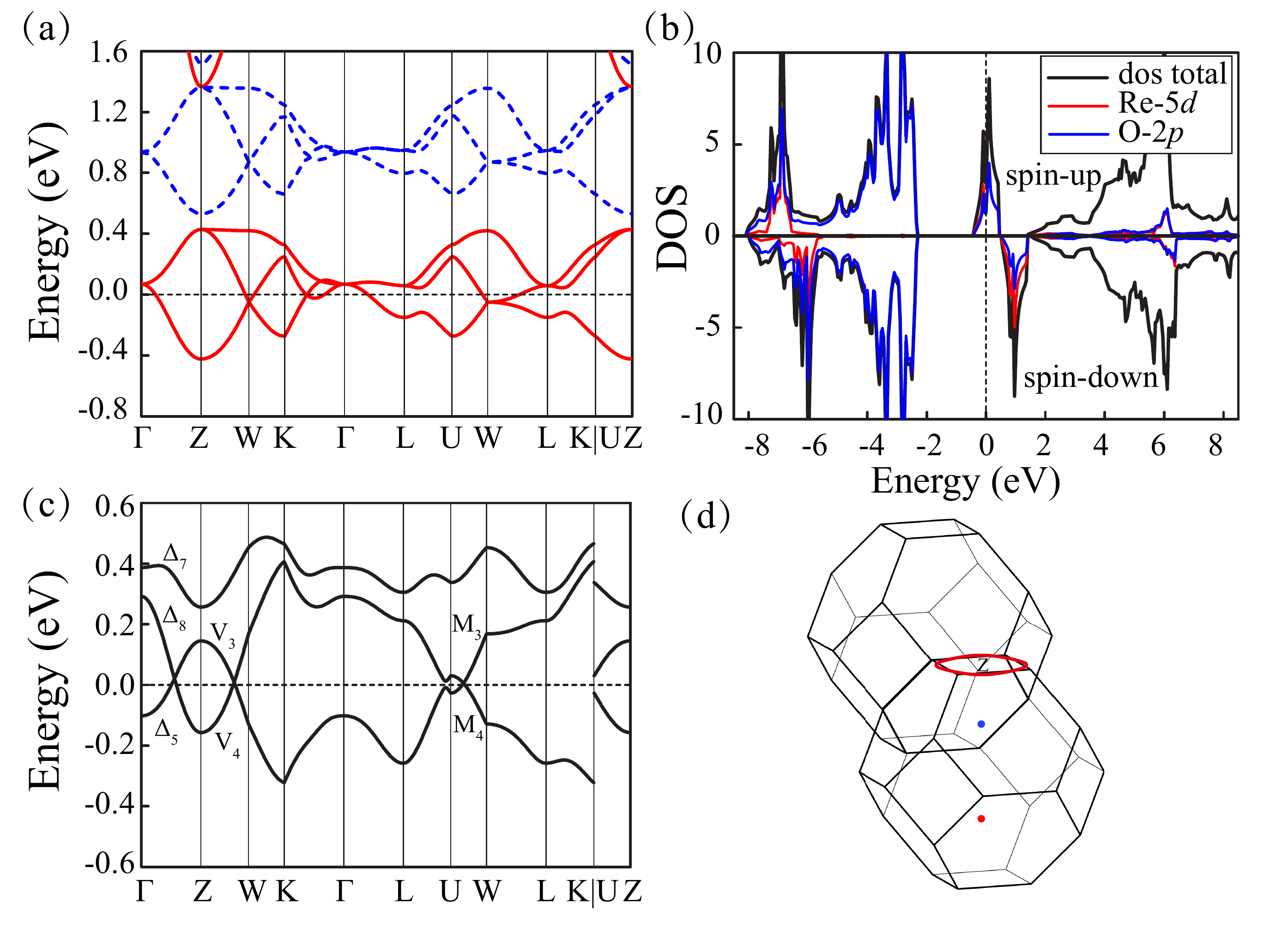}
\caption{\label{bands} (a) Band structure of Ba$_2$CdReO$_6$ without the SOC in the FM ground state. The up (majority) and down (minority) spin bands are denoted by red solid and blue dashed lines, respectively. (b) Total and orbital-resolved partial density of states of Ba$_2$CdReO$_6$. (c) Band structure of Ba$_2$CdReO$_6$ with the SOC included. (d) Distribution of Weyl points and nodal-ring in the bulk BZ. The red/blue dots represent a pair of Weyl points with the opposite chirality, whereas the red line gives the nodal-ring.}
\end{figure}

In the absence of SOC, the internal spin space is decoupled from the lattice space. Here we set the moments of Re atoms aligned along $z$ axis. Figure \ref{bands}(a) shows the electronic band structure of Ba$_2$CdReO$_6$ in the FM ground state, in which the majority spin band is metallic while the minority one is insulating. Focusing on the low-energy bands that are well separated from the others, the linear dispersions with band crossings in the momentum space can be observed at the W point and between the K-$\Gamma$ in the vicinity of the Fermi level. The material is therefore a half-semi-metal with full spin polarization. This can be also clearly seen from the total and orbital-resolved partial density of states (DOS) of Fig. \ref{bands}(b). A trivial band gap of $\sim$2.86\ eV opens in the minority spin channel, whereas the majority one contributes significant states forming a peaklike DOS near the Fermi level, which consist mainly of Re-5$d$ and O-2$p$ orbitals. The material is thus a promising candidate for applications as spintronics devices.


Once the SOC is included in calculations, the two spins couple with each other breaking the spin-rotation symmetry. As we set the moments aligned along $z$ axis, the $M_{x}$, $M_{y}$, $C_{4x}$, and $C_{4y}$ symmetries get broken. However, the $M_{z}$ and $C_{4z}$ symmetries are still preserved. The symmetry of DP Ba$_2$CdReO$_6$  is therefore reduced to $C_{4h}$ magnetic double point group.
Due to the strong SOC, the band structure of DP Ba$_2$CdReO$_6$ has a very distinct change. The band touching points around the Fermi level are mostly gapped. On the $\Gamma$-Z axis, because the symmetry changes from $C_{4v}$ to $C_4$, the double degenerate conduction band splits into two non-degenerate ones, resulting in a crossing of the valence and conduction bands. Since the two bands belong to different irreducible representations ($\Delta_{5}$ and $\Delta_{8}$, respectively), the crossing point is a Weyl point protected by $C_4$ double group symmetry, as shown in the Fig. 2 (c). On the Z-U-W face, the valence and conduction bands also intersect with each other along the Z-W and U-W directions. Further calculations indicate that a nodal-ring is formed on the Z-U-W face due to the band inversion, which is protected by the $M_z$ symmetry, as shown in the Fig. 2(d). As expected, large intrinsic anomalous Hall conductivity, $\sim$400 S/cm, is obtained around the Fermi level in FM DP Ba$_2$CdReO$_6$.

\begin{figure}
\centering
\includegraphics[width=8.5cm]{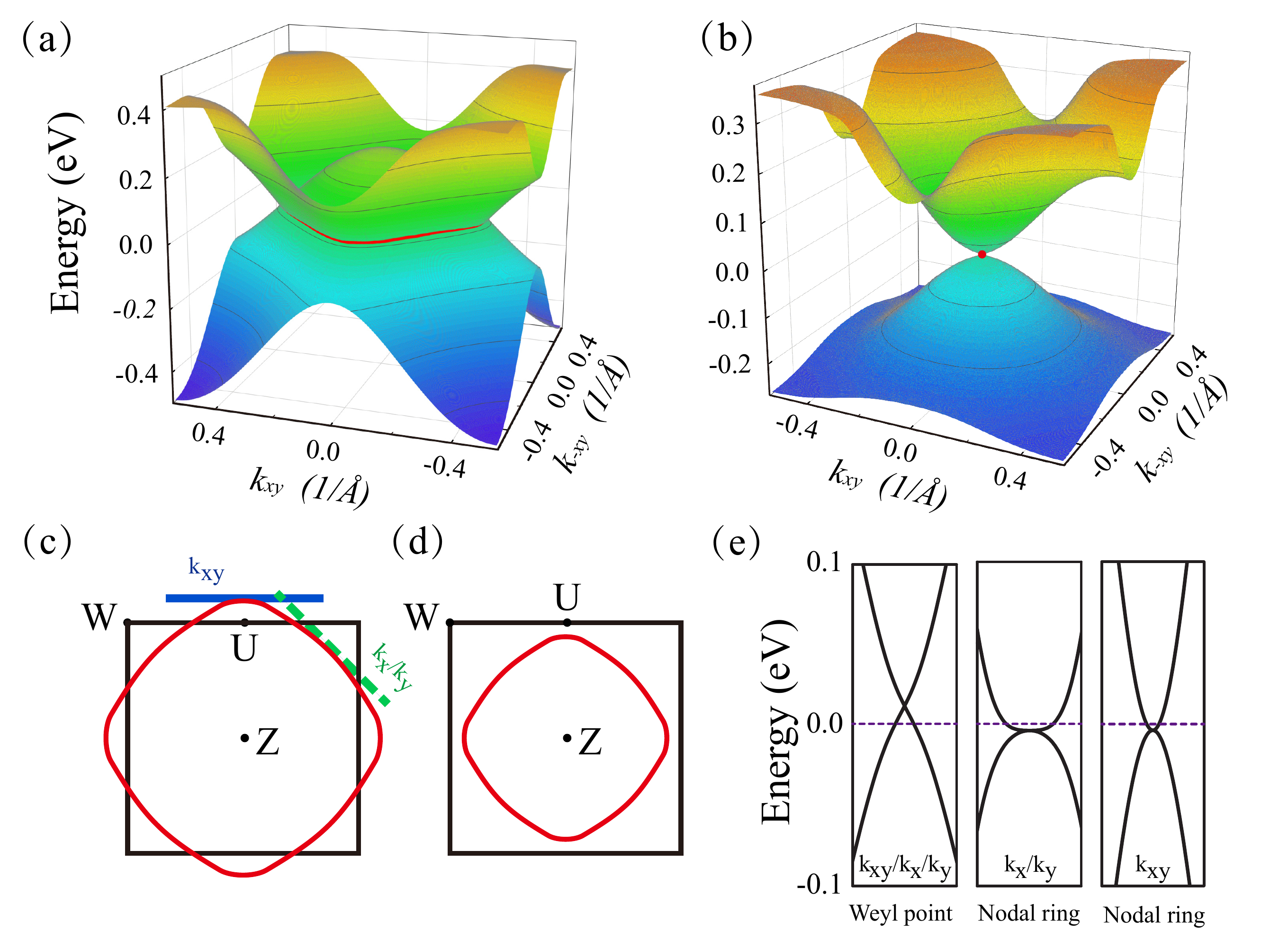}
\caption{\label{plane} The energy dispersion of highest valence and lowest conduction bands of Ba$_2$CdReO$_6$ as functions of $k$, forming nodal-ring and Weyl point in the (a) $k_z$=0.747$\AA^{-1}$ plane and (b) $k_z$=0.413$\AA^{-1}$ plane. Color scale indicates the level of energy. The nodal-ring depicted on the boundary plane of BZ is from calculations with different $U$: (c) $U$=3.0 and (d) $U$=5.5 eV. (e) Left panel shows the band dispersion around the Weyl point along the planar (perpendicular to $k_z$) directions, whereas the middle and right panels give the dispersions along the two tangent directions of nodal-ring as marked in the Fig. \ref{plane}(c).}
\end{figure}

\begin{table}
\centering
\caption{\label{position} The Cartesian coordinates, Chern numbers, and energies (with respect to the Fermi level) of the Weyl points in Ba$_2$CdReO$_6$.}
\begin{tabular}{p{1.0cm}<{\centering} p{2.3cm}<{\centering} p{2.3cm}<{\centering} p{2.0cm}<{\centering}}
\hline\hline
 Point  & Position (\AA$^{-1}$) & Chern number & E-E$_{F}$ (meV)\\
\hline
 W1      & (0, 0, \ 0.413)    &  -1 &  12   \\
 W1${'}$ & (0, 0, -0.413)     & \ 1  & 12   \\
\hline\hline
\end{tabular}
\end{table}

Figures \ref{plane}(a) and \ref{plane}(b) illustrate the energy dispersions of the highest valence and lowest conduction bands of DP Ba$_2$CdReO$_6$ as functions of $k$ in $k_z$=0.747$\AA^{-1}$ and $k_z$=0.413$\AA^{-1}$ planes, respectively. The two bands cross with each other, forming a nodal-ring or Weyl points in the two planes. Around each Weyl point, the bands always disperse linearly, i.e. along the $k_z$ direction ($\Gamma$-Z direction of Fig. \ref{bands}(c)) and $k_{xy}$/$k_x$/$k_y$ direction (left panel of Fig. \ref{plane}(e)). The valence and conduction bands therefore form a perfect Weyl cone in the momentum space. Table \ref{position} lists the Cartesian coordinates, Chern numbers, and relative energies with respect to the Fermi level for the Weyl points, in which the chiral charges are obtained by integrating the Berry curvature of the sphere surrounding each Weyl point in the momentum space \cite{burkov2011weyl}. The relative energies of the  Weyl points in Ba$_2$CdReO$_6$ are about 12 meV above the Fermi level, whereas the nodal-ring states locate slightly below the Fermi level. Interestingly, different with those of Weyl points, the band dispersions of the nodal-ring states are quadratic along $k_x$/$k_y$ or $k_{xy}$ direction, as shown in the middle and right panels of Fig. \ref{plane}(e). This character is similar to the previous observations in nodal-line Heusler semi-metals, which was thought to be one hallmark of topological nodal-line semi-metals \cite{chang2016room}.
Due to the mirror symmetry, the nodal ring has to remain in the $k_z$=$\frac{2\pi}{a}$ plane. An effective Hamiltonian can be constructed to describe the physics as follows \cite{xu2011chern,weng2015topological}:
\begin{eqnarray}
\label{Hkm}
H_{\mathrm{eff}}=\left[\begin{array}{cc}
M& D (k_{z}-\frac{2\pi}{a}) k_{-}^{2} \\
D (k_{z}-\frac{2\pi}{a}) k_{+}^{2} & -M
\end{array}\right] + d_{0}(\boldsymbol{k})
\end{eqnarray}
with eigenvalues $E(\boldsymbol{k})$=$\pm\sqrt{M^{2}+D^{2}(k_{z}-\frac{2\pi}{a})^{2}(k_{x}^{2}+k_{y}^{2})^{2}}\\+d_{0}(\boldsymbol{k})$. Here, $k_{\pm}$=$k_{x} \pm i k_{y}$, $M$=$M_{0}-\beta [k_x^2 +k_y^2+(k_z-\frac{2\pi}{a})^2]$ with $M_{0}$$>$0 and $\beta$$>$0, and $d_{0}(\boldsymbol{k})$ is an energy term in the $k$ space. A flat nodal ring in energy with $k_{x}^{2}$+$k_{y}^{2}$=$\frac{M_{0}}{\beta}$ is obtained for a constant $d_{0}(\boldsymbol{k})$=$d_0$ term, as shown in the Fig. \ref{plane}(a). For a general $k$-point ($m$, $n$) on the nodal ring in the $k_z$=$\frac{2\pi}{a}$ plane, the $k$-cuts along the radial and tangential directions are $k_y$=$\frac{n}{m}k_x$ ($k_x$=0 if $m$=0) and $k_y$=$-\frac{m}{n}k_x$+$\frac{m^2+n^2}{n}$ ($k_x$=$m$ if $n$=0), respectively.
The eigenvalues are therefore $E(\boldsymbol{k})$=$\pm \left(\frac{M_0}{m}k_{x}^2-M_0\right)$+$d_0$ ($E(\boldsymbol{k})$=$\pm \left(\beta k_{y}^2 - M_{0} \right)$+$d_0$ if $m$=0) along the radial direction, leading to a two-bands-inverted picture with linear dispersion around the intersections. Along the tangential direction, the eigenvalues are $E(\boldsymbol{k})$=$\pm\frac{M_0}{n^2}\left(k_{x}-m\right)^2$+$d_0$ ($E(\boldsymbol{k})$=$\pm \beta k_{y}^2$+$d_0$ if $n$=0), which result in two bands with quadratic dispersion touching each other, as shown in the middle and right panels of Fig. \ref{plane}(e). Note that in most topological nodal-line materials, the nodal-line states are stable only without the SOC interaction. They are normally fully gapped or deformed into discrete Weyl points when the SOC is considered \cite{zou2019study}. In DP Ba$_2$CdReO$_6$, however, the nodal-ring is stable in the presence of SOC.

We further examined the effect of on-site Coulomb repulsion $U$, since the bands near the Fermi level are mainly contributed by Re-5$d$ and O-2$p$ orbitals. Different $U$ values ranging from 2.0 eV to 5.5 eV were considered. The bands near the Fermi level are insensitive to the $U$ values used, except that the radius of nodal-ring decreases slightly as $U$ increases, as shown in the Fig. \ref{plane}(c) of $U$=3.0 eV and Fig. \ref{plane}(d) of $U$=5.5 eV. When $U$ is larger than $\sim$5.0 eV, the nodal-ring fully shrinks into the plane of the first BZ. There will be no touching point along U-W (edges of first BZ). However, the Weyl point between $\Gamma$-Z and crossing between Z-W remain.


\begin{figure}
\centering
\includegraphics[width=8.5cm]{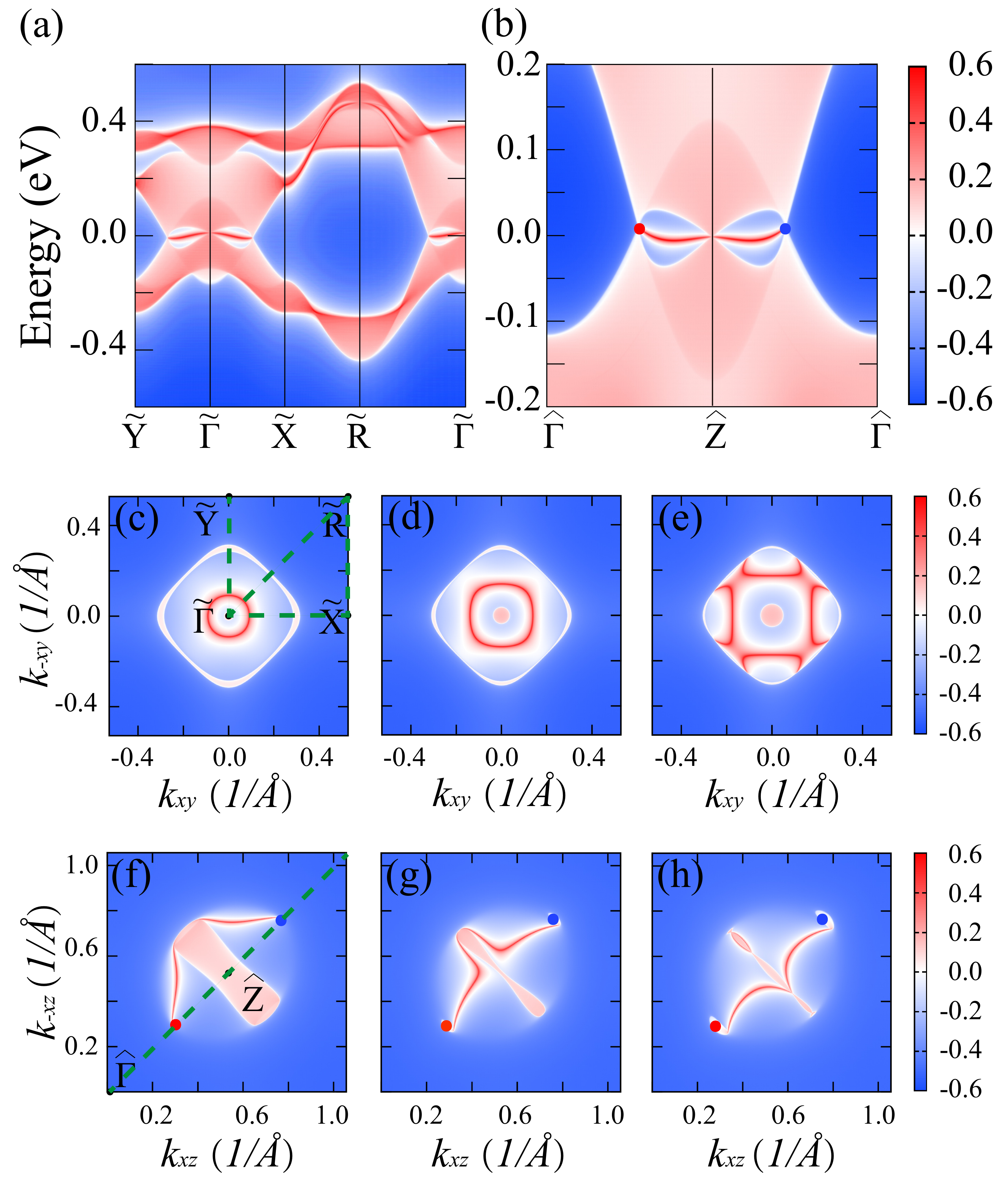}
\caption{\label{surface} The band dispersion of Ba$_2$CdReO$_6$ on the projected (a) (001) surface and (b) (010) surface, along the green dashed lines marked in (c) and (f), respectively. The evolution of corresponding Fermi surfaces and Fermi arcs onto (001) and (010) planes at different Fermi levels (the non-shift one is set to be 0 meV) : (c) 5 meV, (d) 0 meV and (e) -4 meV for (001) surface, and (f) 12 meV, (g) 0 meV, and (h) -4 meV for (010) surface, respectively. The red/blue dots in (f-h) represent the Weyl points  with opposite chirality.}
\end{figure}

The topological surface states of DP Ba$_2$CdReO$_6$ can be obtained with WannierTools software package based on the Green's function method using tight-binding Hamiltonian constructed by the maximally localized Wannier functions \cite{sancho1984quick, marzari2012maximally, mostofi2008wannier90, WU2017}. Figures \ref{surface}(a) and \ref{surface}(b) give the band dispersions projected onto (001) and (010) surfaces, following the high-symmetry $k$-paths marked by the green dashed lines in the Figs. \ref{surface}(c) and \ref{surface}(f), respectively. On the projected (001) surface, topological nontrivial surface states are clearly visible, connecting the two gapless points within $\widetilde{\Gamma}$-$\widetilde{X}$ and $\widetilde{\Gamma}$-$\widetilde{Y}$ and forming a two-dimensional drumhead surface, which is a hallmark of topological nodal-line semimetals \cite{DNL,NL2015}. Moreover, since the width of dispersion is only 20 meV, the drumhead surface states are very flat, generating a large surface density of states. The flat drumhead surface states with full spin polarization could lead to the equal spin-pairing topological superconductivity, such as $p$ or $f$ wave topological superconductors, and is greatly beneficial to spintronics \cite{zhang2020nodal}. As shown in the Figs. \ref{surface}(c-e), the corresponding Fermi surface varies with rigidly shifting the Fermi level $E_F$. It is a small ring centered around $\widetilde{\Gamma}$ when $E_F$=5 meV (the Fermi level of non-shifted band is set to be 0 meV). As $E_F$ decreases, the size of ring increases initially. When the value of $E_F$ is close to the energy of nodal-ring states, the corresponding Fermi surface transforms into a loop centered around the nodal corners, as shown in  Fig. \ref{surface}(e). With a further lowering of $E_F$, the Fermi surface becomes smaller and smaller, leaving only the projected bulk states. Since the Weyl points locate along the high-symmetry \mbox{$Z$-$\Gamma$-$Z$} path, there is no Fermi arc connecting the two Weyl points on the projected (001) surface. However, it is visible on the (010) projected surface. As illustrated in the Figs. \ref{surface}(f-h), there exists a Fermi arc that connects two Weyl points with opposite chirality located between $\hat{\Gamma}$-$\hat{Z}$ of the two adjacent BZs. Part of the Fermi arc merges into projected bulk states. As $E_F$ decreases, the projection of bulk states shrinks. When $E_F$ equals to the energy of nodal-ring, the projected bulk states are mainly composed of nodal-ring states, resulting in coexistence of projected nodal-ring and Fermi arc states on the surface of the material, as shown in the Fig. \ref{surface}(h).

\subsection{Topological half-metallic ferrimagnetism of Ba$_2$FeMoO$_6$}

In this subsection, we reveal the topological aspects of the electronic structures of half-metallic FiM DP compounds with space group symmetry $Fm$$\bar{3}$$m$, such as Ba$_2$FeMoO$_6$, Sr$_2$CrWO$_6$, and Pb$_2$FeMoO$_6$. The arrangement of spins on B and B$'$ transition metal ions is shown in the Fig. 1(f). We find that these materials share similar topological characters and could meet the demands of developing topological quantum spintronics. In the following, we take Ba$_2$FeMoO$_6$ as an example for illustration.

\begin{figure}
\centering
\includegraphics[width=8.5cm]{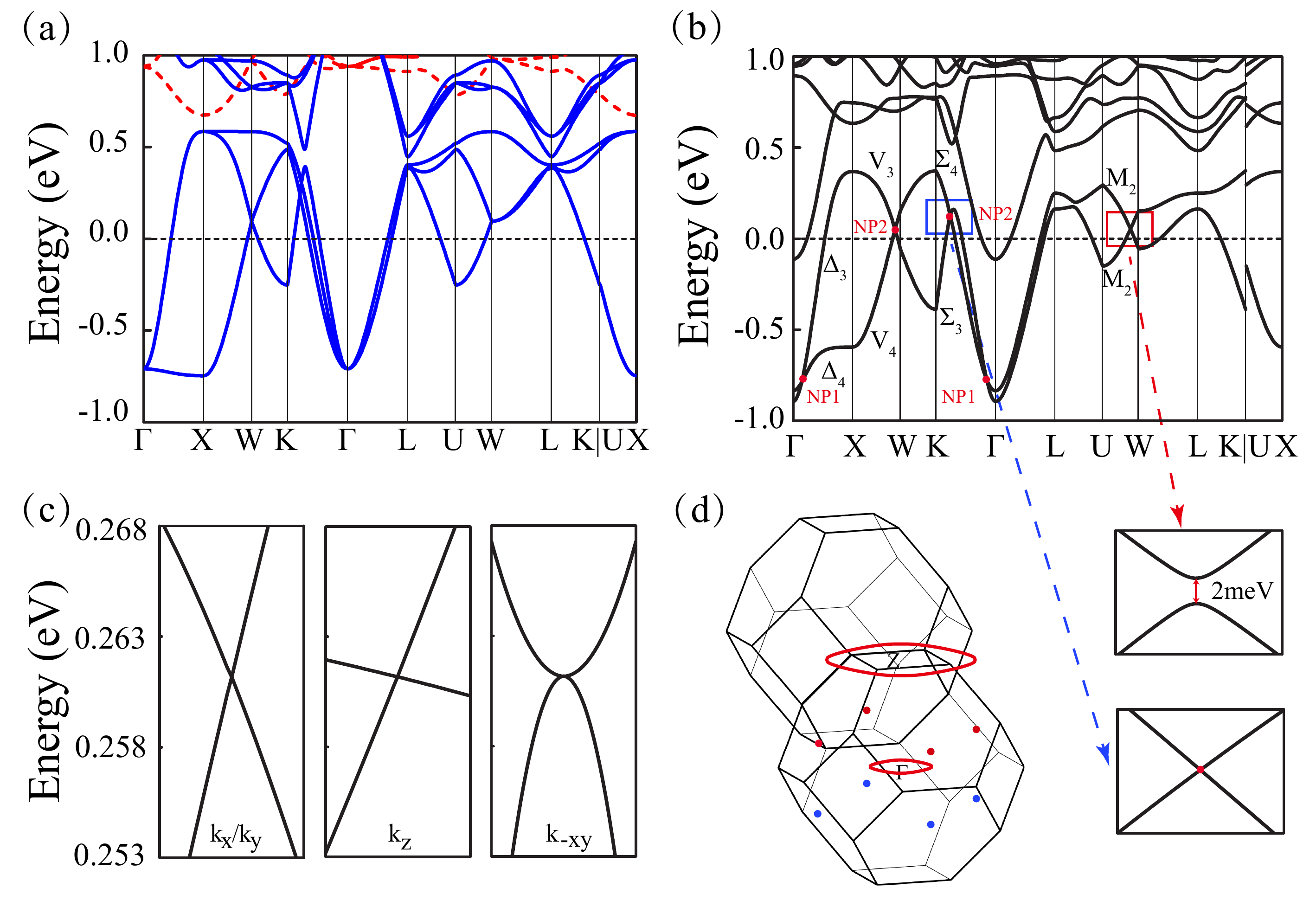}
\caption{\label{FiM-bands} (a) Band structure of Ba$_2$FeMoO$_6$ without the SOC in the FiM ground state. The up (majority) and down (minority) spin bands are denoted by red dashed and blue solid lines, respectively. (b) Band structure of Ba$_2$FeMoO$_6$ with the SOC included. (c) The band dispersions around the Weyl points along different directions. (d) Distribution of Weyl points and nodal-rings in bulk BZ. The eight red/blue dots represent four pair of Weyl points with opposite chirality, whereas the red lines are the nodal-rings.}
\end{figure}

\begin{table}
\centering
\caption{\label{FiM-position} The Cartesian coordinates, Chern numbers, and energies (with respect to the Fermi level) of the Weyl points in Ba$_2$FeMoO$_6$.}
\begin{tabular}{p{0.7cm}<{\centering} p{4.0cm}<{\centering} p{1.5cm}<{\centering} p{1.0cm}<{\centering}}
\hline\hline
\multirow{2}{0.7cm}{Point}   &  Position    & Chern  & E-E$_{F}$  \\
                             & (\AA$^{-1}$) & number &  (meV)     \\

\hline
 W1      & ( -0.410,\ 0.410,  -0.292)   & -1  &  261   \\
 W1${'}$ & ( -0.410,\ 0.410, \ 0.292)   &\ 1  &  261   \\
 W2      & (  0.410,\ 0.410,  -0.292)   & -1  &  261   \\
 W2${'}$ & (\ 0.410,\ 0.410, \ 0.292)   &\ 1  &  261   \\
 W3      & ( -0.410, -0.410,  -0.292)   & -1  &  261   \\
 W3${'}$ & ( -0.410, -0.410, \ 0.292)   &\ 1  &  261   \\
 W4      & (\ 0.410, -0.410,  -0.292)   & -1  &  261   \\
 W4${'}$ & (\ 0.410, -0.410, \ 0.292)   &\ 1  &  261   \\
\hline\hline
\end{tabular}
\end{table}

Figure \ref{FiM-bands}(a) shows the band structure of Ba$_2$FeMoO$_6$ without the SOC in the FiM ground state. The optimized lattice constant after full relaxation of geometries is a=8.14$\AA$, in good agreement with the experimental value \cite{AFMO_Fim}. Different from the case of FM Ba$_2$CdReO$_6$, the minority spin band of FiM Ba$_2$FeMoO$_6$ is metallic while the majority one has an insulating gap of $\sim$2.3 eV. There are three energy bands that cross the Fermi level and contribute states to the Fermi surface, which are mainly composed of Fe-3$d$, Mo-4$d$ and O-2$p$ orbitals. The ordering magnetic moments are mainly around Fe and Mo ions, with opposite spin polarizations $\sim$ 3.9 $\mu_{B}$ and \mbox{-0.5} $\mu_{B}$, respectively. The FiM Ba$_2$FeMoO$_6$ is therefore a half-metal, unlike the semi-half-metallicity of FM Ba$_2$CdReO$_6$. However, linear dispersions with band crossings in momentum space can still be observed in the minority spin channel around the Fermi level. Once the SOC is considered, the band touching points are shifted due to the band splittings, as shown in the Fig. \ref{FiM-bands}(b). Since the $M_{z}$ and $C_{4}$ symmetries are still preserved, there are two fully spin-polarized nodal rings and four pairs of Weyl points formed by the intersections of the highest valence and lowest conduction bands. Their distribution in the BZ is shown in the Fig. \ref{FiM-bands}(d). The nodal-points NP1 at $\sim$ -0.78 eV along the high-symmetry paths $\Gamma$-X and $\Gamma$-K form the smaller nodal-ring around $\Gamma$, while the larger nodal-ring around $Z$ is formed by the points NP2 along the high-symmetry paths W-X and $\Gamma$-K with energies around 0.20 eV. The two nodal-rings are stable against the SOC, due to the protection of symmetry. The two intersecting bands along these paths belong to different irreducible representations, as labeled in the Fig. \ref{FiM-bands}(b). However, the crossing between U-W is opened ($\sim$ 2 meV gap) without any symmetry protection, as illustrated in the red box of Fig. \ref{FiM-bands}(b). We further consider the entire BZ and search for any possible Weyl points, which are not formed along the high symmetry paths of Fig. \ref{FiM-bands}(b). As a result, we find four pairs of Weyl points that are protected by the combined symmetry $\mathcal{CT}$, where $\mathcal{C}$ is the $C_2$ rotation symmetry along [110] ([-110]) direction and $\mathcal{T}$ is the time reversal symmetry. Their Cartesian coordinates, Chern numbers, and relative energies to the Fermi level are listed in Tab. \ref{position}. Note that the band dispersion near the Weyl points does not always behave linearly, i.e. the crossing bands disperse linearly along $k_x$/$k_y$/$k_z$ direction but quadratic along $k_{-xy}$ direction, as shown in the Fig. \ref{FiM-bands}(c).

\begin{figure}
\centering
\includegraphics[width=8.5cm]{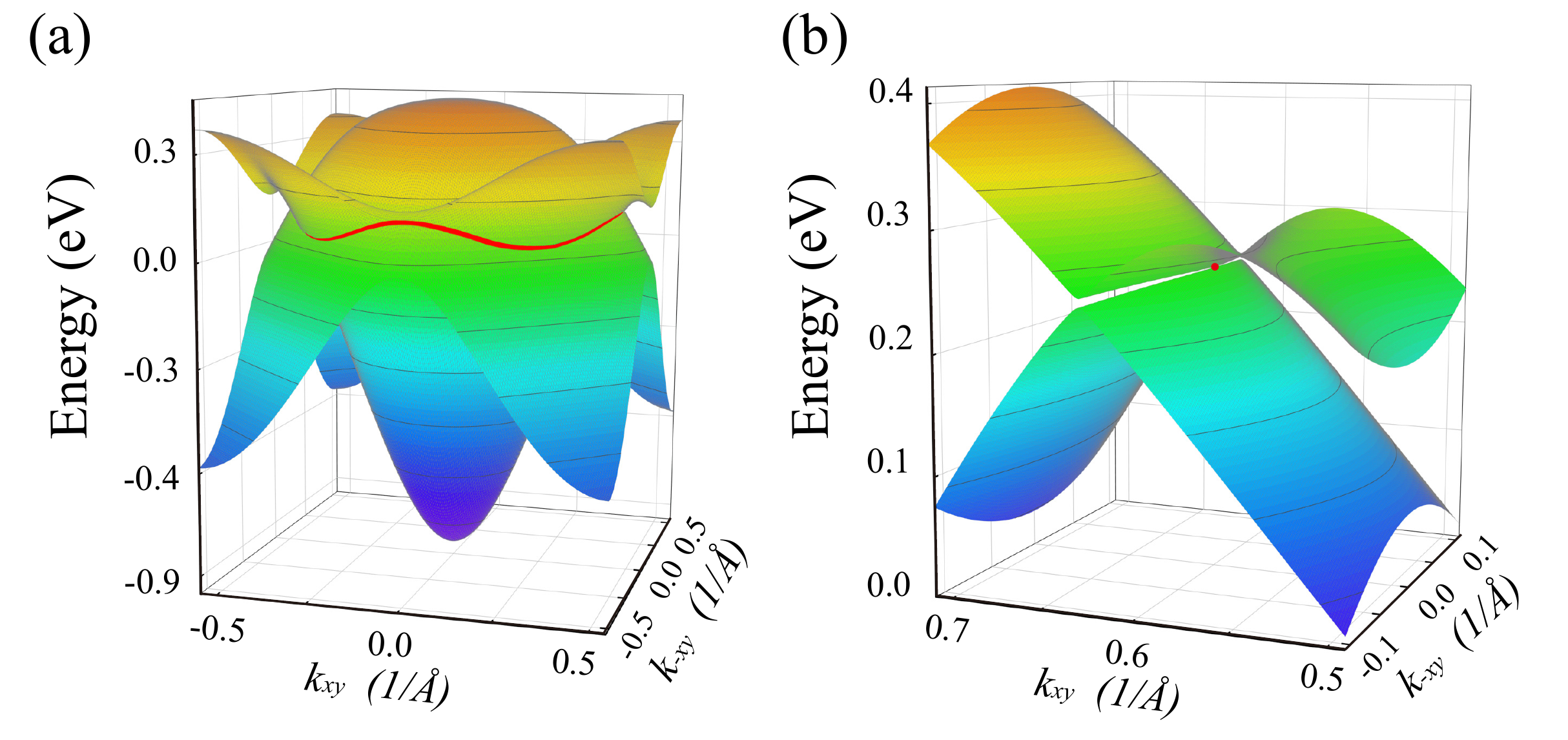}
\caption{\label{FiM-plane} The energy dispersions of highest valence and lowest conduction bands near the Fermi level, which form nodal-ring and Weyl points in the (a) $k_z$=0.772$\AA^{-1}$ plane and (b) $k_z$=0.292$\AA^{-1}$ plane. Color scale indicates the level of energy.}
\end{figure}

The energy dispersions of the highest valence and lowest conduction bands as functions of $k$ in $k_z$=0.772$\AA^{-1}$ and $k_z$=0.292$\AA^{-1}$ planes are shown in the Figs. \ref{FiM-plane}(a) and \ref{FiM-plane}(b), respectively. The two bands touch with each other, forming nodal-rings and Weyl points. The red line marked in the Fig. \ref{FiM-plane}(a) corresponds to the larger nodal-ring around Z. It is shaped like a saddle in the energy space, and has no intersection with the edge plane of the first BZ. As shown by Eq. \ref{Hkm}, the nodal-ring states can be well described by an effective Hamiltonian $H_{\mathrm{eff}}$. The nodal points on the ring do not necessarily have the same energy which is determined by the $d_0(\boldsymbol{k})$ term, for example, the energy of the nodal-point between $\Gamma$-K is higher than that of the point between W-X. However, these nodal-points locate in the same plane of momentum BZ space due to the protection of $M_{z}$ symmetry. The Weyl points instead distribute discretely in BZ, as shown in the Fig. \ref{FiM-plane}(b). As discussed before, the bands near the Weyl point is not always linearly dispersed. The valence band and conduction band therefore do not form a perfect Weyl cone in the momentum space, different from the case of FM Ba$_2$CdReO$_6$.

\begin{figure}
\centering
\includegraphics[width=8.5cm]{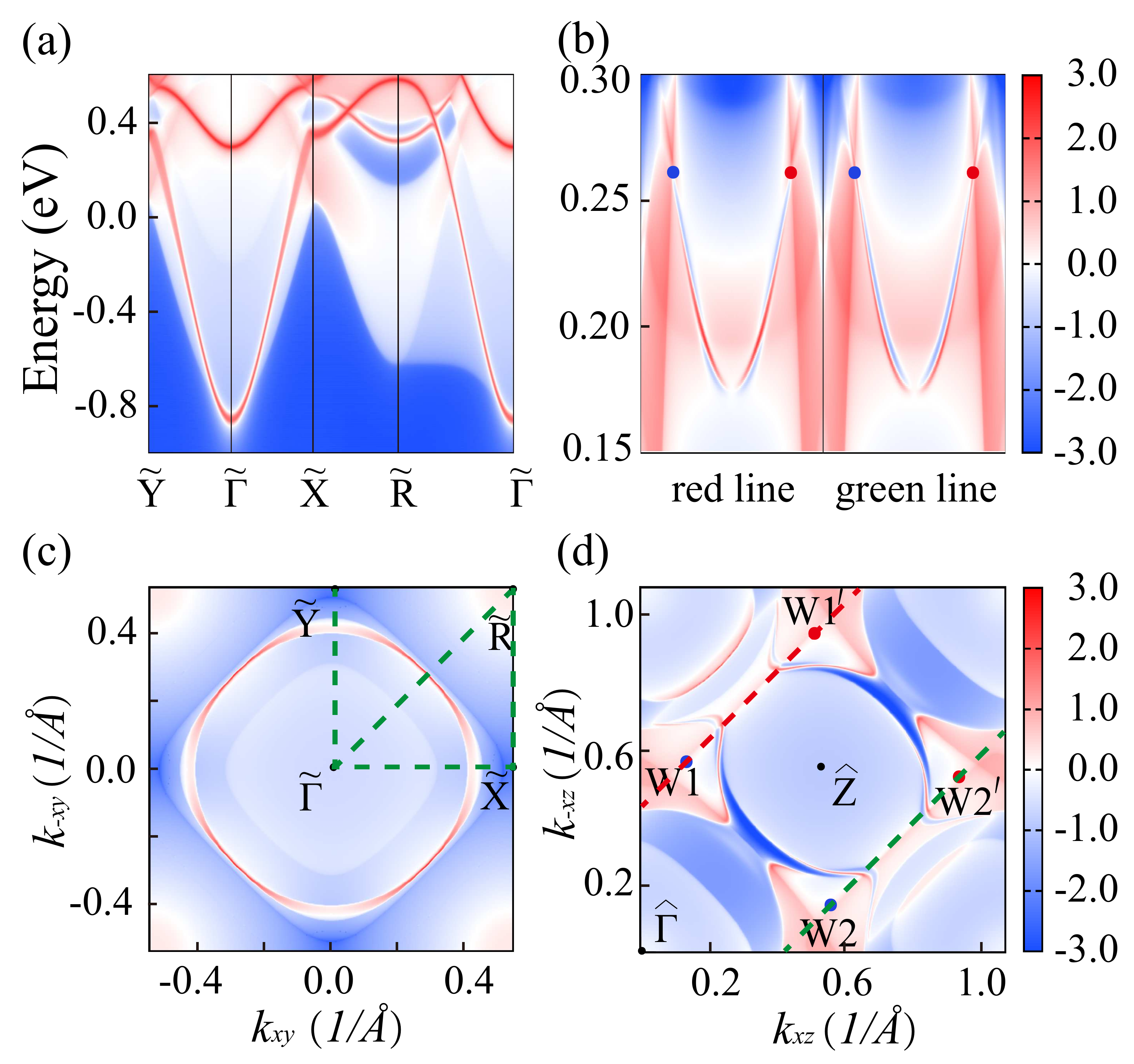}
\caption{\label{FiM-surface} The band dispersions of Ba$_2$FeMoO$_6$ projected onto (a) (001) surface and (b) (010) surface along the dashed lines marked in (c) and (d), respectively. The projected surface states and Fermi arc onto (001) and (010) planes at different rigidly shifted Fermi levels (c) 50 meV and (d) 200 meV are shown, in which the red and blue dots represent the Weyl points with opposite chirality. Here we set the non-shift Fermi level to be 0 meV.}
\end{figure}

Figures \ref{FiM-surface}(a) and \ref{FiM-surface}(b) show the band dispersions of Ba$_2$FeMoO$_6$ projected onto (001) and (010) surfaces, following the $k$-paths marked by the dashed lines in the Figs. \ref{FiM-surface}(c) and \ref{FiM-surface}(d), respectively. On the projected (001) surface, the topological nontrivial surface states are clearly visible, connecting the two nodal points near $\widetilde{Y}$ and $\widetilde{X}$ which are about 0.20 eV above the Fermi level and belong to the larger nodal ring. Since the system is half-metal instead of semi-metal, there is no two-dimensional drumhead-like surface formed. The corresponding Fermi surface projected onto (001) plane is shown in the Fig. \ref{FiM-surface}(c). It is a ring centered around $\widetilde{\Gamma}$, part of whom is merged into the bulk states. The smaller nodal ring is hidden in the bulk states, and hence it is not visible in the projection. As shown in the Fig. \ref{FiM-surface}(b), there exist the surface states directly connecting the pairs of Weyl points with opposite chirality (W1-W1${'}$ and W2-W2${'}$) along the red and green lines in the Fig. \ref{FiM-surface}(d). Two pairs of Weyl points are visible on the projected (010) plane, which overlap with the projections of the other two pairs. Note that the terminations of the corresponding surface Fermi arc are invisible since the projections are plotted slightly away from the energy plane of the Weyl points which are buried into the bulk states, as shown in the Fig. \ref{FiM-surface}(d).

\section{Conclusion}

We predict that both DPs Ba$_2$CdReO$_6$ and Ba$_2$FeMoO$_6$ are topological materials with potential for developing topological quantum spintronic devices. Ba$_2$CdReO$_6$ is an intrinsic FM topological nodal-ring semi-half-metal, in which there coexist a pair of Weyl points along $\Gamma$-$Z$ direction and one nodal ring in the $k_z$=0.747$\AA^{-1}$ plane of BZ. Nearly flat drumhead surface states with full spin polarization are found, which will generate a large surface density of states around the Fermi level. Unlike Ba$_2$CdReO$_6$, FiM Ba$_2$FeMoO$_6$ belongs to topological half-metal, having four pairs of Weyl points and two fully spin-polarized nodal rings in the first BZ. The half-metallicity and topological nontriviality of these DP compounds are robust against the SOC. Our findings therefore provide a wonderful platform for studying the relationship between magnetism and topology, providing a potential high-performance candidate for achieving quantum anomalous Hall effects, topological superconductivity, and device applications.

{\it Note added}: When submitting the manuscript, we noticed that there was a paper just published in PRB \cite{PhysRevB.102.035155}, which reports the similar results for several ferromagnetic double perovskites.

\section{Acknowledgments}

This work was supported by the National Natural Science Foundation of China under Grants No. 11674027 and 11934020. P.-J.G. was supported by China Postdoctoral Science Foundation funded project (Grant No. 2020TQ0347). All calculations were performed at the high performance computing cluster of the Center for Advanced Quantum Studies and Department of Physics, Beijing Normal University, and the National Supercomputer Center in Guangzhou.



\bibliography{reference}

\end{document}